\documentclass[numbers, onecolumn]{rmaa}
\let\gtrsim\relax
\usepackage{paralist}
\usepackage{natbib}
\usepackage{hyperref}
\hypersetup{
    colorlinks=true,
    linkcolor=blue,
    citecolor=blue,
    urlcolor=magenta,
}   
\usepackage{longtable}
\usepackage{array}
\usepackage{amsmath}
\usepackage{booktabs}
\usepackage{amsthm}
\usepackage{caption}
\usepackage{tabularx} 
\usepackage{longtable}
\usepackage{geometry}
\usepackage{graphicx}
\usepackage {float}
\usepackage{supertabular,booktabs}
\usepackage{psfrag,color}
\usepackage[latin1]{inputenc}
\usepackage[dvipsnames]{xcolor}

\title{Investigating open cluster KING 6: Detection of three new variables} 

\author{
  Vaibhav Kumar Pandey,\altaffilmark{1} 
  Arvind K. Dattatrey,\altaffilmark{2}
  Apara Tripathi,\altaffilmark{1}
  R.K.S Yadav,\altaffilmark{2}
  Shantanu Rastogi\altaffilmark{1}}
 
\altaffiltext{1}{Deen Dayal Upadhyaya Gorakhpur University, Gorakhpur, Uttar Pradesh, India, 273009}

\altaffiltext{2}{Aryabhatta Research Institute of Observational Sciences, Manora Peak, Nainital, Uttarakhand,India, 263001}

\shortauthor{Pandey et al.}
\shorttitle{Investigating open cluster King 6}

\listofauthors{vaibhav Kumar Pandey, Apara Tripathi, Arvind Dattatrey, R.K.S Yadav, \& Shatanu astogi}
\indexauthor{Pandey, Vaibhav Kumar}
\indexauthor{Tripathi, Apara}
\indexauthor{Dattatrey, Arvind}
\indexauthor{Yadav, R.K.S.}
\indexauthor{Rastogi, Shantanu}

\abstract{This study presents photometric analysis of the intermediate-age open cluster King 6,  utilizing photometric data in $UBV(RI)_c$ passband and data from the $2MASS$ mission. The Gaia DR3 kinematic data were used to estimate the membership probabilities and $TESS$ data is employed to search for variable stars within the cluster. The cluster's radius is estimated to be 9$^\prime$.0 based on the stellar density profile, while optical and near-infrared color-color diagrams revealed color excesses of \(E(B-V) = 0.58 \pm 0.03\), \(E(J-K) = 0.24 \pm 0.03\), and \(E(V-K) = 1.53 \pm 0.01\) mag. Interstellar extinction law is normal in the direction of the cluster. The cluster$^\prime$s estimated age is $\sim$ 251 Myr and distance is \(724 \pm 5\) pc. The mass function slope was found to be x = $0.57 \pm 0.28$ by considering stars $\geq$ 1 M$_\odot$. Our analysis indicates that the cluster is dynamically relaxed. Furthermore, we identified three new variable stars for the first time in the cluster region using $TESS$ data. These variables belong to the category of slow pulsating B-type variables, with periods of 46.70, 47.92, and 37.56 hours.}

\resumen{}

\addkeyword{Clusters}
\addkeyword{photometric}
\addkeyword{variables}
\addkeyword{stars}
\begin{document}
\maketitle
\section{General}
\label{sec:intro}

Open star clusters serve as fascinating laboratories for exploring our galaxy's stellar evolution and dynamics. They provide insight into star formation processes, enhancing our understanding of the universe \citep{Lada2003}. Open clusters (OCs) consist of stars with similar physical properties, forming simultaneously from the collapse of a molecular cloud \citep{McKee2007}. While member stars share the same age, distances, and chemical compositions, their masses vary  \citep{Joshi2020a}. The study of OCs using astrometric and photometric observations allows us to determine parameters of single star \citep{Dias2021}, partially explaining the interest in these systems. Furthermore, combining astrophysical parameters like distance, metallicity, and age with the kinematic properties of stars within OCs makes them valuable tools for studying formation and evolution of the Galactic disc itself \citep{Cantat2020}. A comprehensive photometric study based on membership determination using proper motions (PMs) and parallaxes helps to understand the stellar and dynamical evolution of the clusters \citep{Tripathi2023}.

King 6  is positioned at ($\alpha_{J2000}$ = 03:27:55.7, $\delta_{J2000}$ = +56:26:38) corresponding to Galactic coordinates $l$ $\sim$ 143$^{\circ}$.36 and $b$ $\sim$ -0$^{\circ}$.07. \citet{Trumpler1930} classified this cluster as II2m based on $V$-band images. \citet{Ruprecht1966} reclassified it in class IV2p. \citet{Ann2002}, using $UBVI_c$ CCD photometry, reported mean reddening $E(B-V)$ = 0.50 $\pm$ 0.10 mag, log(age) = 8.40 $\pm$ 0.10 and distance modulus as $(m - M)_0$ = 9.70 $\pm$ 0.40 mag. \citet{Maciejewski2007} performed a survey in $BV$ wide-field CCD photometry. They estimated core radius as 3$^\prime$.60 $\pm$ 0$^\prime$.40, log(age) as 8.40, distance modulus as 11.17 $\pm$ 0.51 mag, and reddening as 0.53 $\pm$ 0.12 mag suggesting no mass segregation within the cluster. \citet{Piskunov2008} found this cluster to be poorly populated. \citet{Bossini2019} catalogued the age and fundamental parameters of 269 open clusters using data from Gaia DR2 and reported log(age) = 8.58 $\pm$ 0.12, distance modulus 9.54 $\pm$ 0.02, and $A_V$ = 1.06 $\pm$ 0.03 for King 6. \citet{gokmen2023} reported the analysis of King 6 using CCD $UBV$ and Gaia DR3 data. They estimated the colour excess, $E(B-V)$, to be 0.55 $\pm$ 0.03 mag and determined the distance as 723 $\pm$ 34 pc, with an age of 200 $\pm$ 20 Myr. The mass function slope was found to be 1.29 $\pm$ 0.18. Their analysis indicates that the cluster is dynamically relaxed. Table \ref{table5} provides a comparative overview of the parameters we derived alongside previously reported values.

There are differences in the parameter values derived for the cluster in the literature, as shown in Table \ref{table5}. Previous studies on the membership of King 6 did not provide reliable estimates of the cluster members. With the availability of astrometric data from Gaia DR3 and new optical and 2MASS near-infrared data sets, we have revisited the OC King 6 to study their parameters and dynamical status.  Additionally, the extracted parameters of the cluster will be used to enhance the sample of clusters needed for studying Galactic structure and dynamics.

The detection of variable stars provides valuable information for constraining stellar pulsation models and refining theoretical models that predict the characteristics of stars in clusters. Pulsating stars, in particular, offer insights into internal stellar structure and mass. Intermediate-age clusters serve as excellent laboratories for studying short-period variables \citep{Joshi2020b}. Recent work on the Pleiades cluster \citep{bedding2020very} demonstrates the power of combining observations with Gaia \citep{Gaia2016} and $TESS$ (Transiting Exoplanet Survey Satellite) for studying pulsating stars in OCs. The large pulsational frequency separation and the frequency at maximum power can help constrain the stellar evolution, structure and oscillation models, leading to a better characterization of pulsating stars. These advancements have encouraged us to use space-based data to characterize the variables in OCs. In this article, we present the results of our search towards variable stars using observational data of the $TESS$ mission for King 6 and, in turn, reporting their classification.

The paper is structured as follows: Section \ref{Observations and Data Reduction} presents the observations and data reduction. Section \ref{Archived Data} discusses archival data utilized in our analysis. In Section \ref{Structural Parameter of the cluster}, \ref{Proper Motion: Membership}, \ref{Cluster Parameter}, and \ref{Near-IR}, we have discussed the cluster's parameter. In section \ref{Luminosity_mass function}, we have discussed the luminosity and mass function of the cluster followed by mass segregation, relaxation time and tidal radius in Section \ref{segregation}. Section \ref{variable_King6} represents the identification of variable stars in the cluster King 6. Finally, our work is summarized in Section \ref{summary}.

\section{Observations and data Reduction}
\label{Observations and Data Reduction}
\subsection{$UBVRI$ photometric observations}

\begin{figure}[!t]
  \centering
  \includegraphics[width=0.60\textwidth]{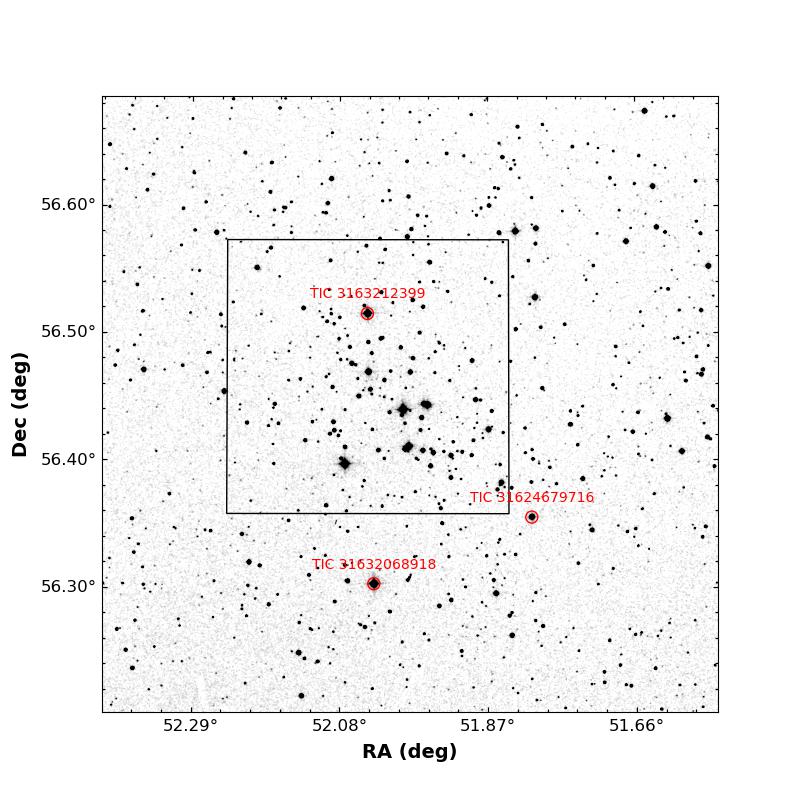}
  \caption{Identification chart for the cluster King 6, taken from SDSS. Rectangular box outlines observed region, while the red circles highlight position of variable stars.}
  \label{finding}
\end{figure}
This study utilizes data collected with the 104-cm Sampurnanand telescope at the Aryabhatta Research Institute of Observational Sciences (ARIES) in Nainital, India. The observations were conducted from December 23$^{rd}$ to 25$^{th}$, 2005. The telescope captured images in multiple bands: $UBV(IR)_C$, using a thinned, back-illuminated CCD camera mounted at the f/13 Cassegrain focus. Figure \ref{finding} presents the finding chart of the cluster, which is taken from the Sloan Digital Sky Survey (SDSS). The rectangular box indicates the region observed using the 104-cm Sampurnanand telescope at Manora Peak, ARIES, Nainital.

The CCD detector has pixels measuring 24 $\mu$m each, arranged in a 2048 $\times$ 2048 grid, providing a field of view on the sky of approximately 13$^\prime$ $\times$ 13$^\prime$, with a pixel resolution of 0$^{\prime\prime}$.38. A binning mode of 2 $\times$ 2 pixels was employed during the observations to enhance the signal-to-noise ratio. The CCD have a readout noise of 5.3 e$^{-}$ and a gain factor of 10 e$^{-}$/ADU. The Observational log is shown in Table \ref{table1}.

\begin{table}\centering
\caption{\bf Description of the optical observation for King 6 and the standard field}
\begin{tabular}{lrrrl}
 \hline
 Cluster/ &Date &Filter\quad&Exp. time $\times$ no.of frames\\
 standard field\\
 \hline
 King\,6&23/24$^{th}$ Dec 2005&$V$& $900\times3$,  $120\times1$ \\
        &&$B$& $240\times2$,  $1200\times3$\\   
        &           &$I$&$60\times1$, $40\times1$\\
         &&  &$100\times2$,  $240\times2$\\
         &&$R$&$480 \times 3$, $60 \times 1$\\
        &            &$U$&$300 \times 2$, $1800\times2$\\ 
 PG1047$+$003&25$^{th}$ Dec 2005&$V$&$120\times4$, $100\times1$\\
                &           &$B$&$200\times5$\\
                &&$I$&$60\times5$\\
                &           &$R$&$60\times5$\\
                &           &$U$&$300\times5$\\
 
 \hline
\end{tabular}
\label{table1}
\end{table}
For image cleaning, bias and twilight flat-field frames were taken. Multiple short and long exposures were collected for the cluster and standard fields in all filters. Calibrating the stellar magnitudes involved observing stars in the standard field PG1047+003. Several \citet{Landolt1992} standard stars were identified within the observed field of PG1047$+$003. The IRAF data reduction package was used for the pre-processing of data frames, which includes bias subtraction, flat fielding, and cosmic ray removal. This process ensures that the data is clean for further analysis.

\subsection{Photometric and astrometric calibration}
Photometry of bias-subtracted and flat-field corrected CCD frames has been carried out using DAOPHOT-II software \citep{Stetson1987, stetson2000}. Quantitative values for the brightness of stars are obtained by aperture photometry and profile-fitting photometry.
Measurements of bright stars, which are saturated in deep exposure frames, have been taken from short-exposure frames. To translate the observed aperture magnitudes to the standard magnitudes, least-square linear regressions are fitted. The equations for calibrating instrumental magnitudes are as follows:\\

\begin{center}
$v$ = $V$ + 4.15 $\pm$ 0.005 - (0.002 $\pm$ 0.004)$(B-V)$ + (0.23 $\pm$ 0.02)$X$
\end{center}

\begin{center}
$b$ = $B$ + 4.57 $\pm$ 0.008 - (0.005 $\pm$ 0.005)$(B-V)$ + (0.38 $\pm$ 0.02)$X$
\end{center}

\begin{center}
$i$ = $I$ + 4.62 $\pm$ 0.010 - (0.020 $\pm$ 0.01)$(V-I)$ + (0.10 $\pm$ 0.02)$X$
\end{center}

\begin{center}
$r$ = $R$ + 4.08 $\pm$ 0.006 - (0.010 $\pm$ 0.004)$(V-R)$ + (0.16 $\pm$ 0.01)$X$
\end{center}

\begin{center}
$u$ = $U$ + 6.79 $\pm$ 0.030 - (0.006 $\pm$ 0.01)$(U-B)$ + (0.60 $\pm$ 0.08)$X$
\end{center}

Here, $V$, $B$, $I$, $R$, and $U$ represent standard magnitudes, and $v$, $b$, $i$, $r$, $u$ refer instrumental aperture magnitudes normalized for 1 sec of exposure time and $X$ is the airmass. The zero points for local standards are evaluated, accounting for the aperture growth curve, difference in exposure times, and atmospheric extinction. The errors associated with the zero points and color coefficients are $\sim$ 0.01 magnitudes.

\begin{figure}[H]\centering
	\includegraphics[width=0.60\textwidth]{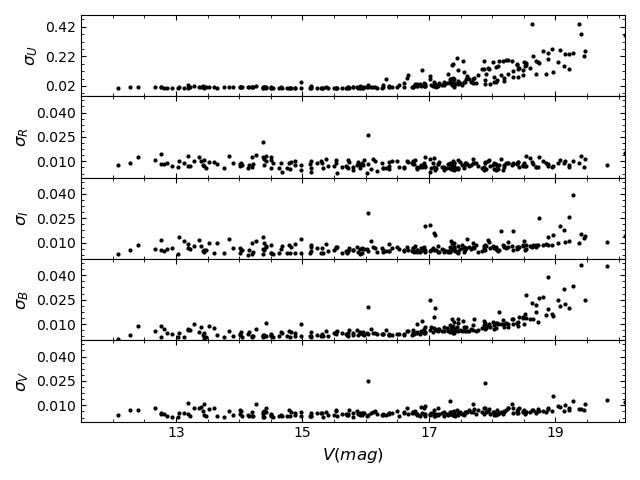}
          \caption{ Photometric errors in $V$, $B$, $I$, $R$, and $U$ against $V$-band magnitude. Errors on the Y-axis represent the internal error as estimated by the DAOPHOT routine.}
      \label{fig_pHotometry_error}
\end{figure}

The internal errors calculated using DAOPHOT are plotted against $V$ magnitudes for the $UBVRI$ filters, as shown in Figure \ref{fig_pHotometry_error}. It shows the average photometric error $\leq$ 0.04 mag across all filters up to a $V$ magnitude of 17 mag. For magnitudes 17 to 20 mag, uncertainity is $\sim$ 0.10 mag whereas for $U$ it is $\sim$ 0.05. Table \ref{table2} analyzes the error trends across filters relative to $V$ magnitude.

\begin{table*}\centering
\setlength{\tabcolsep}{15.0pt}
\caption{Description of observed errors in different filters with $V$ magnitude bins.}
\begin{tabular}{lrrrrrrr}
\hline
$V$ range & $\sigma_V$&$\sigma_B$& $\sigma_I$& $\sigma_R$&$\sigma_U$ \\
\hline
12-13 & 0.01 & 0.01 & 0.01 & 0.01 & 0.01 \\
13-14 & 0.01 & 0.01 & 0.01 & 0.01& 0.02 \\
14-15 & 0.01 & 0.01 & 0.01 & 0.01 & 0.02 \\
15-16 & 0.01 & 0.01 & 0.01 & 0.01 & 0.04 \\
16-17 & 0.01 & 0.02 & 0.03 & 0.01 & 0.10 \\
17-18 & 0.01 & 0.03 & 0.03 & 0.01 & 0.20 \\
18-19 & 0.01 & 0.04 & 0.03 & 0.01 & 0.40 \\
19-20 & 0.01 & 0.10 & 0.04 & 0.01 & 0.50 \\
\hline
\end{tabular}
\label{table2}
\end{table*}

Astrometric solutions are used to get the celestial coordinates of all the stars in $J2000.0$. We use the $CCMAP$ and $CCTRAN$ tasks of $IRAF$ to obtain the celestial coordinates of the detected stars. The sources detected in different wavebands are merged using a matching radius of $\sim$ 2 $^{\prime\prime}$. In total, 3028 stars are detected in all five bands.
 \begin{figure}[!t]\centering
\centering
	\includegraphics[width=0.65\textwidth]{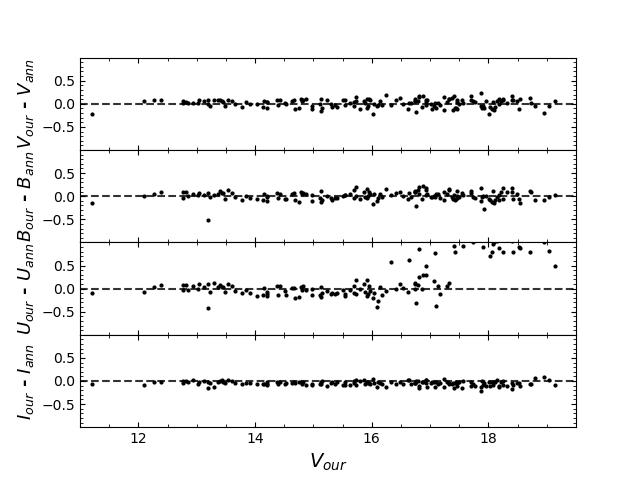}
          \caption{Comparison of the present photometry with  \citet{Ann2002} in $U, B, V$ and $I_c$ filters against $V$ magnitude. The dotted lines are plotted at zero value.}
      \label{fig_verification}
\end{figure}
\subsection{Comparison with the previous photometry}
The CCD $UBVI$ photometry was previously done by \citet{Ann2002}. To verify our photometry, we have performed the cross-identification of the stars with the photometry by \citet{Ann2002}. We have cross-identified 158 stars. We estimated the difference in magnitude and plotted it against $V$ magnitude as shown in Figure \ref{fig_verification}. The mean of the differences and standard deviation (i.e mean $\pm$ $\sigma$)
for $V, B, I_c$ and $U$ magnitude is found to be 0.02 $\pm$ 0.09, 0.01 $\pm$ 0.09, -0.05 $\pm$ 0.05, and 0.04 $\pm$ 0.64 mag,  respectively.

\begin{table*}[!t]
\centering
\setlength{\tabcolsep}{15.0pt} 
\caption{Description of uncertainties in proper motions and magnitudes against Gaia G magnitude bins.}
\begin{tabular}{lrrrrrr}
 \hline
 G(mag) &$\sigma_{\mu_\alpha cos{\delta}}$ &$\sigma_{\mu_{\delta}}$ &$\sigma_{G}$ &$\sigma_{G_{RP}}$ &$\sigma_{G_{BP}}$ \\
 \hline
 11-12 & 0.020 & 0.020 & 0.002 & 0.003 & 0.004 \\
 12-13 & 0.020 & 0.020 & 0.002 & 0.003 & 0.004 \\
 13-14 & 0.030 & 0.030 & 0.002 & 0.004 & 0.004 \\
 14-15 & 0.040 & 0.040 & 0.003 & 0.005 & 0.005\\
 15-16 & 0.050 & 0.050 & 0.003 & 0.006 & 0.005\\
 16-17 & 0.070 & 0.070 & 0.003 & 0.010 & 0.005 \\
 17-18 & 0.114 & 0.110 & 0.003 & 0.030 & 0.007 \\
 18-19 & 0.198 & 0.196 & 0.004 & 0.070 & 0.013 \\
 19-20 & 0.384 & 0.385 & 0.005 & 0.120 & 0.024 \\
 \hline
\end{tabular}
\label{table3}
\end{table*}

\section{Archival Data}
\label{Archived Data}
\subsection{Gaia DR3}
\begin{figure}[!t]\centering
\centering
	\includegraphics[width=0.65\textwidth]{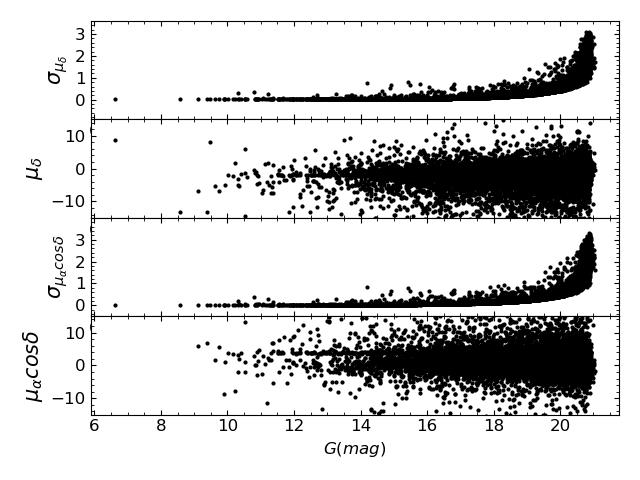}
          \caption{ Proper motions in the direction of RA and DEC and its error are plotted against $G$ mag.}
      \label{fig_gaia}
\end{figure}

The Gaia DR3 data is used for astrometric study and to determine the structural parameters of cluster. It provides celestial positions and $G$ band magnitudes for a vast dataset of approximately 1.8 billion sources, with magnitude measurements extending up to 21 mag. Additionally, Gaia DR3 provides valuable parallax, proper motion, and colour information ($G_{BP}-G_{RP}$ ) for a subset of this data set, specifically 1.5 billion sources. The uncertainties in parallax values are $\sim$ 0.02 - 0.03 milli arcsecond (mas) for sources at $G \leq 15 $ mag and $\sim$ 0.07 mas for sources with $G \sim 17 $ mag. We have collected data for King 6 within a radius of 30$^{\prime}$. The proper motions of the stars in the cluster and corresponding errors are graphically represented against $G$ magnitude in Figure \ref{fig_gaia}. The lowest panel displays two prominent structures. Uncertainties in the corresponding proper motion components are less than 0.04 mas up to 15$^{th}$ magnitudes; $\leq$ 0.10 mas from 15$^{th}$ to 18$^{th}$ mag and less than 0.40 mas up to 20$^{th}$ magnitudes. Detailed analysis of uncertainties in proper motion along Right Ascension (RA) and Declination (DEC) and Gaia magnitudes ($G$, $G_{BP}$, $G_{RP}$) is shown in Table \ref{table3}.

\subsection{The Near-infrared Data}
We used archival near-infrared photometric data from the Two Micron All-Sky Survey (2MASS) \citep{skrutskie2006two}), which provides photometry in the $J$ (1.25 $\mu$m), $H$ (1.65 $\mu$m), and $K_s$ (2.17 $\mu$m) filters. The data have limiting magnitudes of 15.80, 15.10, and 14.30 in the $J$, $H,$ and $K_s$ bands, respectively, with a signal-to-noise ratio (S/N) greater than 10. Our optical data were cross-correlated with the 2MASS photometric catalogue, resulting in the identification of 233 common stars within a matching radius of 2$^{\prime\prime}$.

\subsection{TESS (Transiting Exoplanet Survey Satellite) data}
 We have utilized $TESS$ data. which features a large plate scale of 21$^{\prime\prime}$ per pixel. This resolution often causes the TESS light curves to capture the combined light from multiple stars \citep{higgins2023localizing}.  In our investigation of the variable stars in our cluster, we identified three isolated stars after matching the $TESS$ data with our optical observations.

\begin{figure}[!t]\centering 
 \centering
 \includegraphics[width=0.95\textwidth]{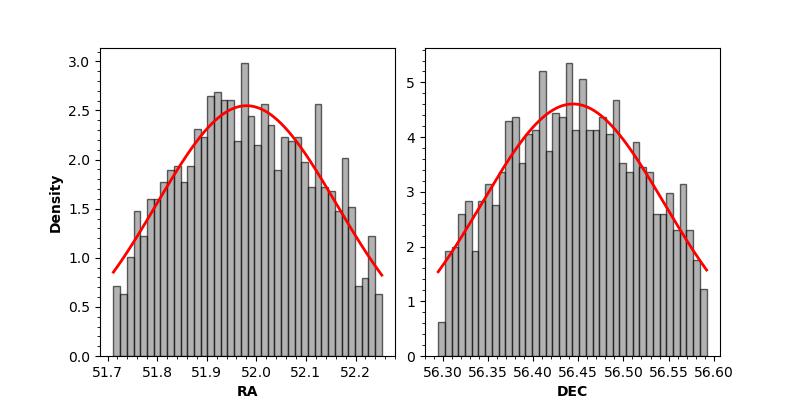}
 \caption{Histogram in RA and DEC of the stars. Solid red line represents the fitted Gaussian function.}
 \label{fig_center}
\end{figure}

\section{Structural Parameter of the cluster: Cluster extent}
\label{Structural Parameter of the cluster}
The size of a star cluster plays a crucial role in studying its dynamic evolution, which involves how the cluster changes over time due to internal and external forces. However, determining the precise size of a cluster can be challenging due to its irregular shape, making it difficult to pinpoint the exact centre and outer boundary. To estimate the centre of the cluster, we employed a histogram technique. We plotted the number of stars against their celestial coordinates (RA and DEC) as shown in Figure \ref{fig_center}. The data for this analysis comes from Gaia DR3, specifically for stars with a $G$-band magnitude uncertainty $\leq$ 0.10 magnitudes and located within a 30$^\prime$.0 radius around the cluster centre listed in Table \ref{table5}. Our analysis reveals cluster centre at approximately (03:27:55.26, +56:26:39.07) for King 6, which is roughly 15$^{\prime\prime}$.0 away from the centre estimated by \citet{Ann2002}. 

Defining the cluster radius, which indicates the point at which the internal gravitational influence becomes negligible, is a crucial parameter. We achieve this by analyzing the stellar density profile. To conduct a quantitative analysis, we created concentric annular regions (ring-shaped areas) centred on the estimated cluster centre. We then counted the number of stars within each annular ring (width of 0$^{\prime}$.50) and divided this count by the corresponding area of the ring to calculate the number density of stars in that region. This method allows us to examine how the density of stars varies with distance from the cluster center, ultimately helping us determine the effective radius of the cluster. To estimate this radius, we fitted the surface density profile described by \citet{King1962} to the radial distribution of stars. This fit is accomplished using a non-linear least squares routine that takes the error as a weighting factor. The radial density profile (RDP) can be represented as follows:

\begin{equation}
\rho(r) = \frac{\rho_0}{1+(\frac{r}{r_c})^2} + \rho_b \ 
\end{equation}

where $\rho_0$ is central density, $r_c$ is core radius of the cluster and $\rho_b$ is the background density.

\begin{figure*}[!t]\centering
\centering
	\includegraphics[width=0.65\textwidth]{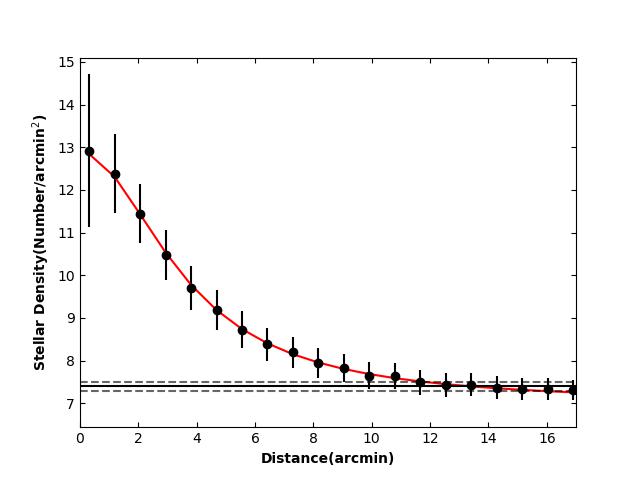}
          \caption{Radial density profile of the cluster King 6. Red solid curve represents best fit of  \citet{King1962}. Solid horizontal line represents the background density, wheras dashed lines represent the errors.}
      \label{fig_RDP}
\end{figure*}

Figure \ref{fig_RDP} illustrates the best-fit solution for the density distribution of King 6, along with the associated uncertainties. As evident from the graphs, the radial density profile for the clusters exhibits a decline followed by a flattening behaviour at approximately 9$^\prime$.0 for the cluster. Based on this observation, we have adopted 9$^\prime$.0 as the cluster radius, which is slightly lower than the values previously reported by \citet{Maciejewski2007} and \citet{gokmen2023}, as shown in Table \ref{table5}. For King 6, $\rho_0$ = 5.88 stars/arcmin$^{2}$, $r_c$ = 3.58 arcmin and $\rho_b$ = 7.00 stars/arcmin$^{2}$. The value of core radius $r_c$ is comparable within error to the values reported by \citet{gokmen2023} (see Table \ref{table5}).

\section{Determination of membership probability}
\label{Proper Motion: Membership}
\begin{figure}[!t]\centering
\centering
	\includegraphics[width=0.80\textwidth]{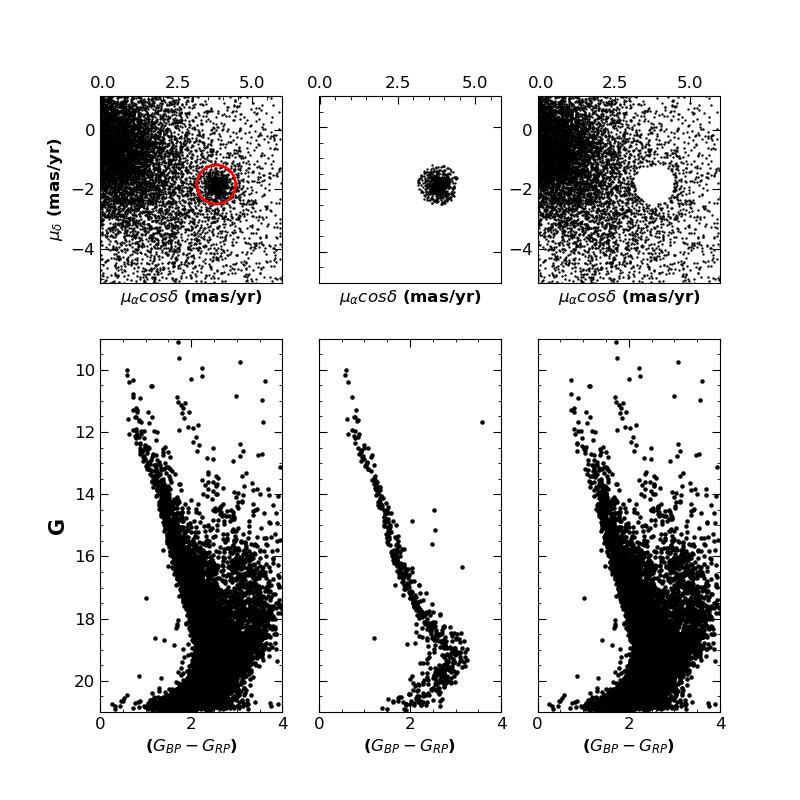}
          \caption{Initial cluster member separation using proper motion. Top
panels show the VPDs and bottom panels show the respective CMDs for total, cluster and field region.}
      \label{fig_vdp}
\end{figure}

 A significant challenge in studying OCs is the presence of unrelated stars along the line of sight, known as field star contamination. Distinguishing these field stars from the cluster's actual members is crucial for accurately estimating the cluster's physical and dynamical properties \citep{Angelo2019, Angelo2020}. An effective method for separating members from field stars is the vector point diagram (VPD) based on proper motions \citep{Dias2018a}. The unprecedented astrometric precision of the Gaia DR3 catalogue has significantly improved the reliability of membership determination based on kinematic data (data related to stellar motion) \citep{Castro-Girnard2018,Castro-Girnard2019}; \citep{LiuandPang2019}. Therefore, we used data from Gaia DR3 for our kinematic analysis, member star identification, and cluster distance calculations. Figure \ref{fig_vdp} illustrates the VPDs constructed using the proper motion of stars in RA and Dec for the cluster. Here, the cluster stars are separated from the field stars, with the cluster members encircled in red as shown in the upper panel of the figure, and the cluster members are separated from the field stars shown by CMD in the lower panel of Figure \ref{fig_vdp}. This distinction is possible because cluster stars exhibit a more concentrated distribution in their proper motions than field stars. The centre of the red circle is determined using the maximum density method used by \citet{Joshi2020a}.

The estimated value of the centre of cluster proper motion distribution is found to be ($\mu_{\alpha}\cos{\delta}$, $\mu_{\delta}$) = 3.82 mas yr$^{-1}$, -1.90 mas yr$^{-1}$ for King 6. We identified stars within the red circle as probable member stars. We found a total of 575 probable member stars for the cluster. To quantify the membership of stars, we calculated the membership probabilities. We estimated membership probabilities using a statistical approach based on PMs of stars as described in previous studies (\citet{Sanders1971}; \citet{Michalska2019}; \citet{Pandey2020}).
 
\begin{figure}[!t]\centering
\centering
	\includegraphics[width=0.70\textwidth]{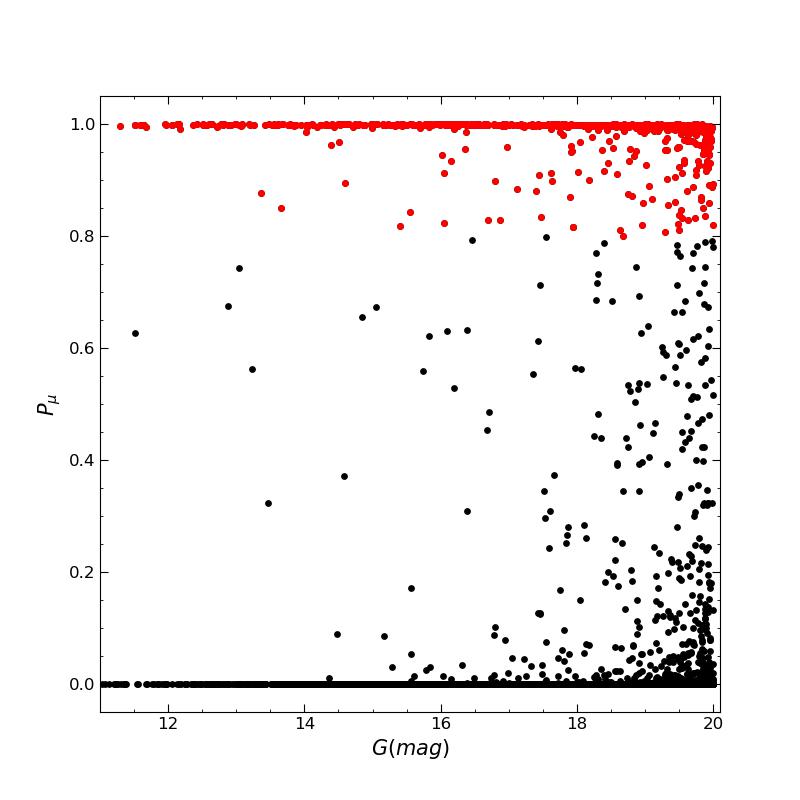}
          \caption{The Gaia membership probability is plotted against $G$ magnitude. The red dot represents the star with a probability of $P_\mu\ge$ 0.80, and the black dots represent the star with a probability of $P_\mu <$ 0.80.}
      \label{fig_membership}
\end{figure}

Assuming a distance of 0.75 kpc and a radial velocity dispersion of 1 km s$^{-1}$ for open clusters \citep{Girard1989}, a dispersion ($\sigma_c$) of $\sim$0.28 mas yr$^{-1}$ in the proper motion of the cluster can be obtained. We calculated $\mu_{xf} = 0.15$ mas yr$^{-1}$, $\mu_{yf} = -0.74$ mas yr$^{-1}$, $\sigma_{xf} = 5.76$ mas yr$^{-1}$, and $\sigma_{yf} = 5.53$ mas yr$^{-1}$ for the field stars. These values are used to construct the frequency distributions of the cluster stars ($\phi^{\nu}_c$) and the field stars ($\phi^{\nu}_f$) using the equation given in \citet{Yadav2013}. The value of the membership probability for the $i^{th}$ star is calculated using the equation given below:

\begin{equation}
P_\mu(i) = \frac{n_c\times\phi^\nu_c(i)}{n_c\times\phi^\nu_c(i)+n_f\times\phi^\nu_f(i)}
\end{equation} 
where $n_c$ and $n_f$ are the normalized numbers of probable cluster members and field stars, respectively. 
We have found 606 stars with $P_\mu$ $>$ 0.80 for the cluster. In Figure \ref{fig_membership}, we have plotted the membership probability against $G$ magnitude. In this figure, we can see the separation between cluster members and field stars toward brighter part, which supports the effectiveness of this technique. A high membership probability extends to $G$ $\sim$ 20  mag, whereas the probability gradually decreases toward the fainter limits. 
 We matched our $UBVRI$ catalogue with the Gaia catalogue using a matching radius of 2$^{\prime\prime}$.0. We found 242 optical counterparts of Gaia sources with membership probability information. We found 94 stars within the cluster radius (9$^\prime$) having membership probability $P_\mu$ $\geq$ 0.80.

\section{Cluster parameters}
\label{Cluster Parameter}
\subsection{$(U-B)$ vs $(B-V)$ diagram}
\label{reddening_(U-B}
\begin{figure}[!t]\centering
 \centering
	\includegraphics[width=0.65\textwidth]{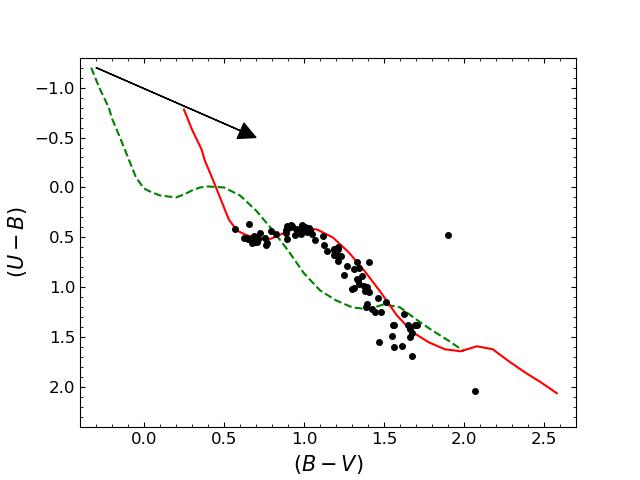}
          \caption{The colour-colour diagram for target region. Dashed line represents intrinsic ZAMS, while the solid curve represents best-fit ZAMS line. Arrow illustrates the reddening vector.}
      \label{fig_reddening}
\end{figure}

 The reddening, $E(B-V)$, can be determined by fitting intrinsic zero-age main-sequence (ZAMS) on the $(U-B)/(B-V)$ colour-colour diagram \citep{Phelps1994}. The black points denote member stars present within a 9$^{\prime}$.0 radius and with probability $P_\mu$ $>$ 0.80. The best fit of the intrinsic ZAMS  was achieved by shifting $E(B-V)$ along the reddening vector as shown in Figure \ref{fig_reddening}. We adopted the slope of the reddening vector $E(U-B)/E(B-V)$ = 0.72 and fitted the ZAMS with the solar metallicity, i.e., $Z$ = 0.02 \citep{Schmidt1982}. The best fit was achieved for the reddening value of 0.58 $\pm$ 0.03 mag. Our estimated value of $E(B-V)$ = 0.58 $\pm$ 0.03 mag is comparable to the values estimated by \citet{Maciejewski2007},  \citet{gokmen2023}, and  \citet{Ann2002} (see Table \ref{table5}).

\subsection{Total-to selective extinction value}
\begin{figure}[!t]\centering
 \centering
    \includegraphics[width=0.65\textwidth]{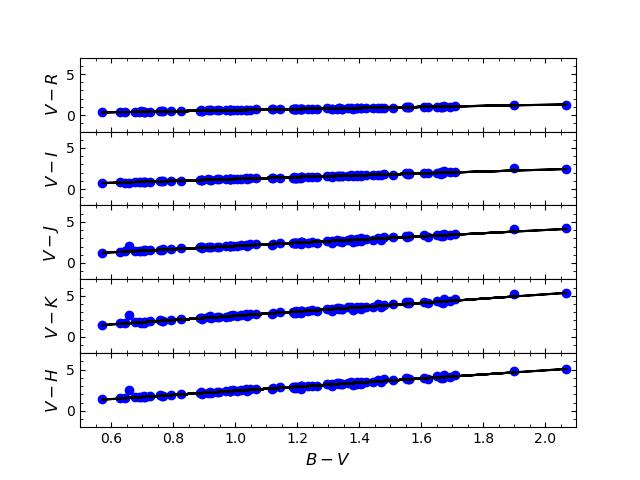}
    \caption{\textbf{$(V-\lambda)/(B-V)$ colour-colour diagram is plotted for stars within cluster radius, with membership probability $P_\mu$ $>$ 0.80. Continuous line represents the slopes.}}
      \label{fig_CCD}
\end{figure}

Reddening is an important parameter for a star cluster since it can significantly affect the determination of other fundamental parameters. Colour-colour diagrams are essential for understanding the extinction law. The emitted photons of cluster stars are scattered and absorbed in the interstellar medium by dust particles, which leads to the deviation of colours from their intrinsic values. The normal Galactic reddening law may not accurately describe the reddening observed along the line of sight to many star clusters \citep{Sneden1978}. To examine the nature of the reddening law,  two colour diagrams, $(V-\lambda)/(B-V)$,  are employed as suggested by \citet{Chini1990}, where $\lambda$ represents all filters other than $V$. Here, we have examined the reddening law for the cluster by drawing two colour diagrams, as shown in Figure \ref{fig_CCD}. Since the stellar colour values emerge to show the linear relationship, a linear equation is applied to calculate the slope (m$_{cluster}$).  We have calculated the total-to-selective extinction using the relation given by \citet{Neckel1981}:

\begin{equation}
    R_{cluster} =  \frac{m_{cluster}}{m_{normal}}\times R_{normal}
\end{equation}

where $R_{normal}$ is the normal value of total-to-selective extinction taken to be 3.10, and m$_{normal}$ and m$_{cluster}$ are the slopes of a linear fit to two colour diagrams for normal extinction and extinction in the direction of clusters, respectively. The values of m$_{normal}$ and m$_{cluster}$ are given in Table \ref{table4}. The average $R_{cluster}$  (=2.78 $\pm$ 0.30) is approximately equal to the normal value. Thus, the reddening law is normal in the direction of the cluster.

\begin{table*}\centering
\setlength{\tabcolsep}{10.0pt}
\caption{The slopes of the ($V-\lambda)/(B-V)$ diagrams for the member stars and corresponding normal values.}
\begin{tabular}{lrrrrr} 
   \hline
    &$\frac{(V-R)}{(B-V)}$&$\frac{(V-I)}{(B-V)}$&  $\frac{(V-J)}{(B-V)}$&$\frac{(V-K)}{(B-V)}$&  $\frac{(V-H)}{(B-V)}$\\
    \hline
   Normal &$0.80$&$1.70$&$2.23$&$2.72$&$2.42$\\
   King 6&$0.66\pm0.001$&$1.34\pm0.001$&$1.94\pm0..001$&$2.61\pm0.001$&$2.49\pm0.001$\\
   $R_{cluster}$&2.56&2.44&2.70&2.97&3.19\\
   \hline
\end{tabular}
\label{table4}
\end{table*}

\subsection{Distance and Age Determination}
\begin{figure}[H]\centering
\centering
	\includegraphics[width=10cm,height=8cm]{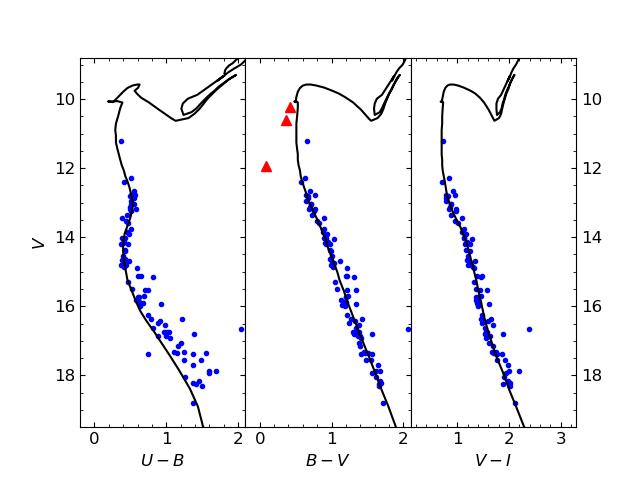}
          \caption{The colour-magnitude diagrams for the member stars of the cluster with  $P_\mu$ $\geq$ 0.80. The red triangles represent positions of variable stars using data from \citet{Stassun2019}.}
      \label{fig_CMD_0ptical}
\end{figure}

A CMD has been used to estimate the cluster's age and distance. $V$ versus $(U-B)$, $V$ versus $(B-V)$, and $V$ versus $(V-I)$ CMDs of the member stars (represented by blue dots) within the cluster radius 9$^\prime$.0 and the $P_\mu$ $\geq$ 0.80 are shown in Fig \ref{fig_CMD_0ptical}. The CMDs of the target cluster show a well-defined narrow main sequence (MS). MS extends from $V$ $\sim$ $11.2-18.8$ mag. The theoretical isochrone of solar metallicity $Z$ = 0.02 taken from \citet{Bertelli1994} have been overplotted on the CMDs and represented by solid lines. The reddening values derived in Section \ref{reddening_(U-B} were used for the isochrone fitting. The best fit of the theoretical isochrone gives the value of distance modulus as $(V-M_v)$ = 11.10 $\pm$ 0.40 mag, giving the distance to be 724 $\pm$ 5 pc. Our estimated distance value is comparable with the values derived by \citet{Maciejewski2007} and  \citet{gokmen2023} (see Table \ref{table5}).  

\begin{figure}[!t]\centering
\centering
	\includegraphics[width=0.65\textwidth]{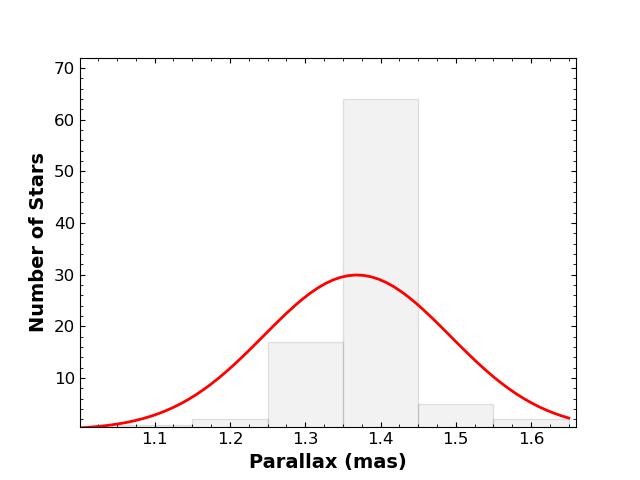}
          \caption{ Histogram in parallax (0.10 mas bins) for the cluster region. Red curve represents the Gaussian function fitting.}
      \label{fig_histo_plx}
\end{figure}

We have also used parallax of the stars to obtain the distance of the cluster. In Figure \ref{fig_histo_plx}, we plotted the histogram of 0.10 mas parallax bin using the star within the cluster radius with the probability membership $P_\mu$ $\geq$ 0.80. A Gaussian function is fitted on the histogram. We found a mean parallax of 1.36 $\pm$ 0.10 mas. Using this mean parallax, we estimated the distance as 735 $\pm$ 10 pc, which closely agrees with the value derived using CMDs.

The age of the cluster has been determined by fitting theoretical stellar evolutionary isochrones of different ages on the CMDs ($V/(U-B)$, $V/(B-V)$, and $V/(V-I)$). The isochrone fitting is done considering the bluest envelope of the CMDs, as shown in Figure \ref{fig_CMD_0ptical}. After correcting for reddening, the best fit is obtained for an age of log(age) = 8.40 ($\sim$ 251 Myr), which is in good agreement with the previously estimated values by \citet{Ann2002}; \citet{gokmen2023} and \citet{Maciejewski2007}, listed in Table \ref{table5}. 
\section{Intersteller extinction in near-IR}
\label{Near-IR}

\subsection{$(V-K)$, $(J-K)$ diagram}
We have combined $2MASS$ $JHK_s$ data with the optical data to determine interstellar extinction in near-IR. The $K_s$ magnitude is converted into $K$ magnitude using the methods given by \citet{Persson1998}. The $(J-K)$ verses $(V-K)$ colour-colour diagram for the cluster members is shown in the left panel of Figure \ref{fig_UVBRIJHK}. The solid line in the figure shows the ZAMS given by \citet{Bertelli1994}. This ZAMS provides $E(J-K)$ = 0.24 $\pm$ 0.03 mag and $E(V-K)$ = 1.50 $\pm$ 0.01 mag for the cluster. The ratio $\frac{E(J-K)}{E(V-K)}$ is $\sim$ 0.16 $\pm$ 0.03, which is in good agreement with the normal interstellar extinction value 0.19 given by \citet{Cardelli1989}. This scatter is due to the large error in the $JHK$ data.

\subsection{$(B-V)$ vs $(J-K)$ diagram}
We have plotted $(B-V)$ versus $(J-K)$ colour-colour diagram for the cluster, as shown in the middle panel of the Fig \ref{fig_UVBRIJHK}. We have used the ZAMS given by \citet{Bertelli1994} to know the relationship between these two colours. The colour excess $E(B-V)$ and $E(J-K)$ is found to be 0.58 and 0.24 mag. This results in a ratio of $\frac{E(B-V)}{E(J-K)} \sim$ 0.43 mag, which is slightly lower than the value of 0.56 reported in the literature \citep{Cardelli1989}. 

\subsection{$(J-H)$ vs $(J-K)$ diagram}
\begin{figure}[!t]\centering
 \centering
	\includegraphics[width=0.65\textwidth]{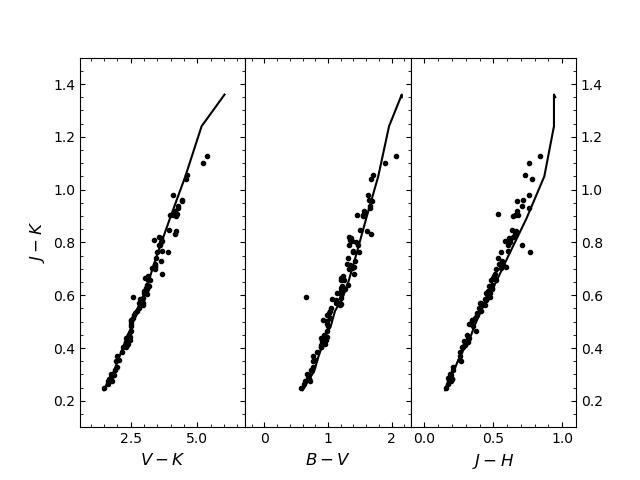}
          \caption{Colour-colour diagram  for the stars within the cluster radius and $P_\mu$ $\geq$ 0.8.} Solid line is the ZAMS taken from \citet{Bertelli1994}.
      \label{fig_UVBRIJHK}
\end{figure}

The $(J-H)$ vs $(J-K)$ colour-colour diagram for the cluster is shown in the right panel of Figure \ref{fig_UVBRIJHK}. To determine the relation between these two colours we have utilized ZAMS given by \citet{Bertelli1994} as indicated by the solid line in the right panel of the figure. We have found that $E(J-H)$ = 0.15 $\pm$ 0.02 and $E(J-K)$ = 0.24 $\pm$ 0.03. We have calculated the value of the ratio $\frac{E(J-H)}{E(J-K)}$ = 0.64 $\pm$ 0.06 which is in agreement within 2$\sigma$ the normal value 0.55 as quoted by \citet{Cardelli1989}. 

\subsection{Age and distance of the cluster using $2MASS$ data}
\begin{figure}[!t]\centering
\centering
	\includegraphics[width=0.65\textwidth]{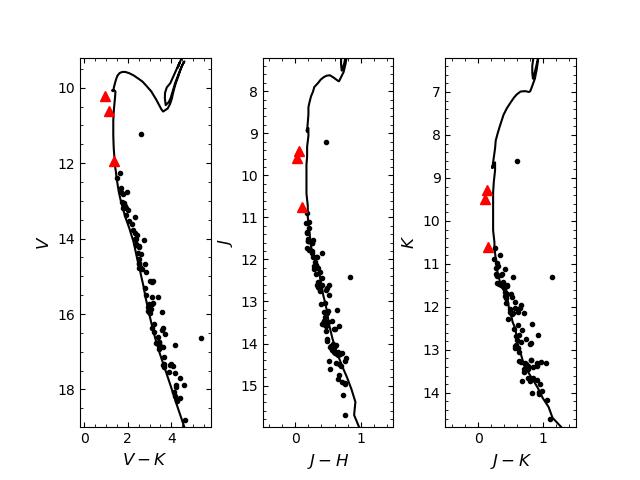}
          \caption{Colour-magnitude diagram for the stars within cluster radius and  $P_\mu$ $\geq$ 0.80. Triangular red dots represent the position of variable stars.}
      \label{fig_CMD_JHK}
\end{figure}

Using 2MASS data, we re-determined the distance and age of the cluster. We have plotted $V$ versus $(V-K)$ in the left panel, $J$ versus $(J-H)$ in the middle panel, and $K$ versus $(J-K)$ in the right panel in Figure \ref{fig_CMD_JHK}. We have used the theoretical isochrone given by \citet{Bertelli1994} for $Z=0.02$ of log(age) = 8.40, which has been overplotted in the CMDs. We have found the apparent distance modulus of 11.10 $\pm$ 0.08, 9.82 $\pm$ 0.08, and 9.57 $\pm$ 0.08 mag using the CMDs respectively shown in Figure \ref{fig_CMD_JHK}. This distance modulus leads to a distance of 733 $\pm$ 7 pc. The age and the distance of the cluster derived using 2MASS data agree with those derived using optical data. 

\section{Luminosity and mass function}
\label{Luminosity_mass function}
\subsection{Luminosity Function}
\begin{figure}[!t]\centering 
\centering
	\includegraphics[width=0.65\textwidth]{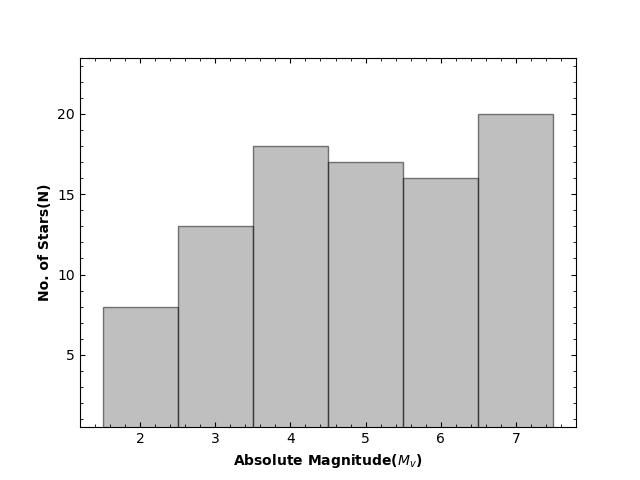}
	\caption{Luminosity function of the cluster region.}
\label{luminosity_plot}
\end{figure}

We examined the brightness distribution of stars within the cluster. To achieve this, we considered the stars with the probability $P_\mu$ $\geq$ 0.80 and within the cluster radius. We considered their apparent brightness ($V$ magnitude) and converted it to their true brightness (absolute magnitude) by accounting for the cluster's distance and a correction factor ($A_V$). We grouped stars in brightness (bins of 1.00 magnitude) to ensure enough stars in each group for reliable analysis. Finally, by plotting the number of stars in each brightness bin, we created a luminosity function(LF) for the cluster, shown in Figure \ref{luminosity_plot}. Interestingly, this distribution shows an increasing trend up to $M_v$ $\sim$ 4 mag and a dip at $\sim$ 5.50 mag for the cluster.

\subsection{Mass Function}
The mass function (MF), i.e., the frequency distribution of stellar masses, is a fundamental parameter in studying star formation and evolution in the cluster. It represents the distribution of stellar masses per unit volume in a star formation event, and knowledge of MF is very effective in determining the subsequent evolution of the cluster. The MF can be derived using the linear relation :
\begin{equation}
log(dN/dM) = -(1+x)\times log M + constant
\end{equation}

\begin{figure}[!t]\centering 
\centering
	\includegraphics[width=0.65\textwidth]{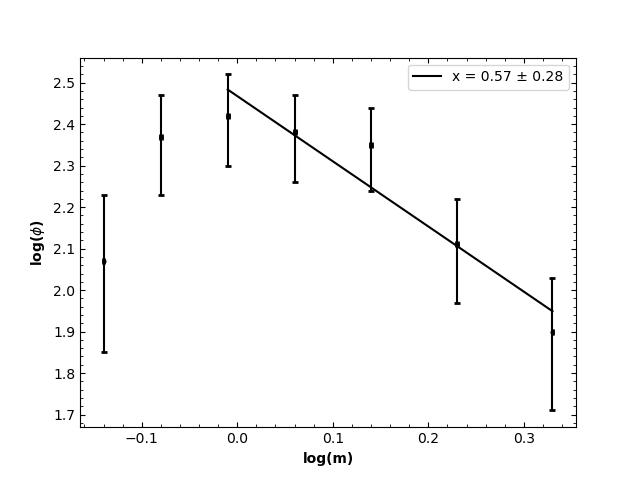}
	\caption{Mass Function for cluster under study  derived using \citet{Bertelli1994} isochrone. The error bars represent $\frac{1}{\sqrt{N}}$}
\label{mass_function}
\end{figure}

where $dN$ represents the number of stars per mass bin $dM$ with the central mass $M$, and $x$ is the slope of the MF. To convert the LF to the MF, we used the theoretical evolutionary model given by \citet{Bertelli1994}, and the resulting MF is shown in Figure \ref{mass_function}. This figure shows a turn $\sim$ 1 $M_{\odot}$. To derive the mass function slope, we consider the stars to have a mass $\geq$ 1 $M_{\odot}$. The derived slope of the MF is $x = 0.57\pm0.28$, which is less than the value 1.35 given by \citet{Salpeter1955} for Solar neighbourhood stars. The calculated mass function slope is less than the mass function slope 1.29 $\pm$ 0.18 determined by \citet{gokmen2023} within the mass range 0.58 $\geq$ M/M$_\odot$$\geq$ 3.59. 

\section{Mass segregation, relaxation time and tidal radius of the cluster}
\label{segregation}
\subsection{Mass segregation}
\begin{figure}[H]\centering
\centering
	\includegraphics[width=0.65\textwidth]{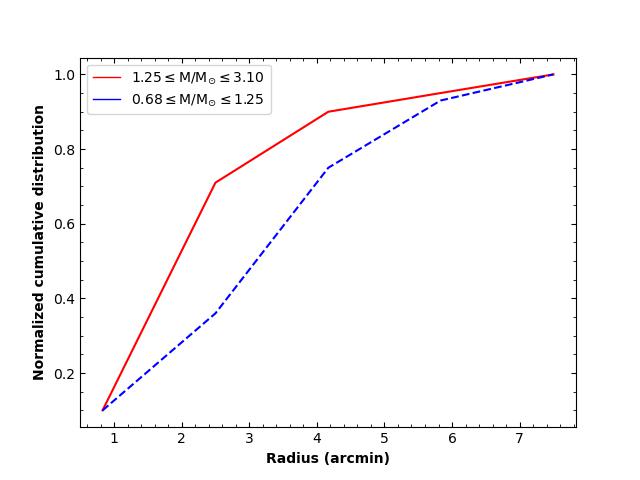}
	\caption{The cumulative radial distribution of member stars in low-mass ($< 1.25
M_\odot$) and relatively higher-mass ($\geq 1.25 M_\odot$) range .}
\label{mass_segregation}
\end{figure}

To study the mass segregation of the cluster, we plotted the cumulative radial distribution of the stars for different masses as shown in Figure \ref{mass_segregation}. To do so, we have divided the main sequence stars in the two mass ranges, $3.10\leq M/M_{\odot}\leq1.25$ and $1.25\leq M/M_{\odot}\leq0.68$ for the cluster. To obtain the mass segregation effect, we have selected the probable member based on the membership probability and CMD of the cluster. The figure exhibits the mass segregation effect in the sense that massive stars gradually sink towards the cluster centre rather than the fainter star. To check whether this mass distribution represents the same kind of distribution, we have performed the Kolmogorov-Smirnov(K-S) test. This test indicates that the confidence level of the mass segregation effect is 98 $\%$ for the cluster.

\subsection{The relaxation time}
The relaxation time ($T_E$) is the time scale in which the cluster loses all traces of its initial condition. It is the peculiar time scale for the cluster to reach some extent of energy equipartition. The relaxation time is given by \citet{Spitzer1971}:
\begin{equation}
    T_E = \frac{8.9\times10^5(NR_h^3)^{1/2}}{<m>^{1/2}log(0.4N)}
\end{equation}

where $N$ is the number of cluster members with probability ($P_\mu$ $\geq$ 0.80) and within radius 9$^\prime$.00, $R_h$ is the half-mass radius of the cluster and $<m>$ is the mean mass of the cluster stars \citep{Spitzer1971}. In our case, the value of $<m>$ is found to be 1 $M_\odot$ The value of the $R_h$ is assumed to be the half of the radius derived by us in Section \ref{Structural Parameter of the cluster}. Using the above relation, we estimated the cluster's dynamical relaxation time $T_E$ to be 6.0 Myr. A comparison of the cluster's relaxation time with the cluster's age indicates that the relaxation time is smaller than the cluster's age. Therefore, we conclude that the cluster is dynamically relaxed. 

\subsection{Tidal radius}
The tidal radius of the cluster is the extent to which the gravitational influence of the Galaxy is equal to that of the gravitational influence caused by the cluster core. It can be determined using the following procedure.

The Galactic Mass $M_G$ inside a Galactocentric Radius $R_G$ is given by \citep{genzel1987physical}
\begin{equation}
    M_G = 2\times10^8M_\odot(\frac{R_G}{30pc})^{1.2}
\end{equation}
The value of $M_G$ is found to be 1.83$\times$10$^{11}$M$_\odot$. Using the formula by \citet{Kim_2000}, the tidal radius of the cluster can be obtained as 
\begin{equation}
    R_t = (\frac{M_c}{2M_G})^{1/3}\times R_G
\end{equation}
where $R_t$ and $M_c$ are the cluster's tidal radius and total mass, respectively. The mass of the cluster is calculated by considering the overall mass function slope derived in this article within the mass range 0.68-3.10$M_{\odot}$. Thus, the derived tidal radius of the cluster is found to be 3.18 pc. A comparative analysis of the fundamental parameters of King 6, as derived in this study and reported in the literature, is presented in Table \ref{table5}.
\begin{table}
\centering
    \caption{Comparison of our derived fundamental parameters of King 6 with the literature values}     
    \label{table5}
      \setlength{\tabcolsep}{8.0pt}
    \begin{tabular}{ccc}   
      \toprule
      Parameters & Numerical Values&Reference\\
      \midrule
    (RA, DEC) (deg) &(51.98 $\pm$ 0.18, 56.44 $\pm$ 0.10)& Present study\\
                    &51.98, 56.44& \citep{Ann2002}\\
    $(\mu_\alpha \cos\delta$, $\mu_\delta$) (mas yr$^{-1}$)&$(-3.82, -1.90)$& Present study\\
    Cluster Radius (arcmin) & 9$^\prime$.00& Present study\\
                            &10$^\prime$.90& \citep{Maciejewski2007}\\
                            &10$^\prime$.00& \citep{gokmen2023}\\
    Cluster Radius(pc) & 1.98 & Present study\\
    Tidal Radius(pc) & 3.18& Present study\\
    $\rho_0$( stars/arcmin$^{2}$) & $5.88 \pm 0.87$ & Present tudy\\
                                   & $2.28 \pm 0.24$& \citep{gokmen2023}\\
    $r_c$(arcmin) & $3.58 \pm 0.49$  & Present study\\
                & $4.68\pm 1.07$ & \citep{gokmen2023}\\
    $\rho_b$(stars/arcmin$^{2}$) & $7.00 \pm 0.06$ & Present study\\
                                & $5.12 \pm 0.16$ & \citep{gokmen2023}\\
    log(age) & $8.40$ & Present study\\
             & $8.40 \pm 0.10$ & \citep{Ann2002}\\
             & $8.40$ & \citep{Maciejewski2007}\\
            & $8.58 \pm 0.12$ & \citep{Bossini2019}\\
            & $8.29$ & \citep{Dias2021}\\
            & $8.30$ & \citep{gokmen2023}\\
    Distance (pc) & $724 \pm 5 $ & Present study\\
                & $426$ & \citep{Ann2002}\\
                  & $800\pm 270$ & \citet{Maciejewski2007}\\
                  & $497$ & \citep{Bossini2019}\\
                  & $704$ & \citep{Dias2021}\\
                  & $723 \pm 34$ & \citep{gokmen2023}\\
    $E(B-V)$(mag) & $0.58 \pm 0.03$ & Present study\\
             & $0.50 \pm 0.10$ & \citep{Ann2002}\\
             & $0.53 \pm 0.12$ & \citep{Maciejewski2007}\\
             & $0.34$ & \citep{Bossini2019}\\
             & $0.59$ & \citep{Dias2021}\\
             & $0.55 \pm 0.03$ & \citep{gokmen2023}\\
    $E(J-K)$(mag) & 0.24 $\pm$ 0.03 & Present study\\
    $E(V-K)$(mag) & 1.50 $\pm$ 0.01 & Present study\\
    $R_{cluster}$ & 2.78 $\pm$ 0.30 & Present study\\
    $A_v$  & 1.80 & Present study\\
           & $1.596 \pm 0.09$ & \citep{gokmen2023}\\ 
    Mass function slope (x) & 0.57 $\pm$ 0.28 & Present study\\
    Relaxation Time(Myr) & 6.0 & Present study\\
    \bottomrule  
    \end{tabular}    
\end{table} 

\section{Variable stars in the cluster King 6}
\label{variable_King6}
The variable stars are the objects that change their brightness over time. This variation provides critical insight into the physical processes governing stars. In the present study, we used the $TESS$ data to search the variable stars in the cluster region. We found three variable stars (TIC 31624679716, TIC 31632068918, and TIC 3163212399) that were isolated in the SDSS image. 
These variables are plotted in the CMDs shown in Figure \ref{fig_CMD_0ptical} and Figure \ref{fig_CMD_JHK}. We have used the light curve provided by $TESS$ to analyse these variables. The periodicity of these variables is determined using the Lomb-scargle algorithm provided by \citet{Scargle1982}, \citet{lomb1976} for unevenly spaced light curves. The effective temperature (T$_{eff}$) taken from SIMBAD is 11055, 10510, and 10510 K for TIC 31624679716, TIC 31632068918, and TIC 3163212399 stars, respectively. The value of temperature suggests that they are B-type stars. The light curves of each variable are analysed below.

\subsection{TIC 31624679716}
\begin{figure}\centering
\centering
	\includegraphics[width=0.65\textwidth]{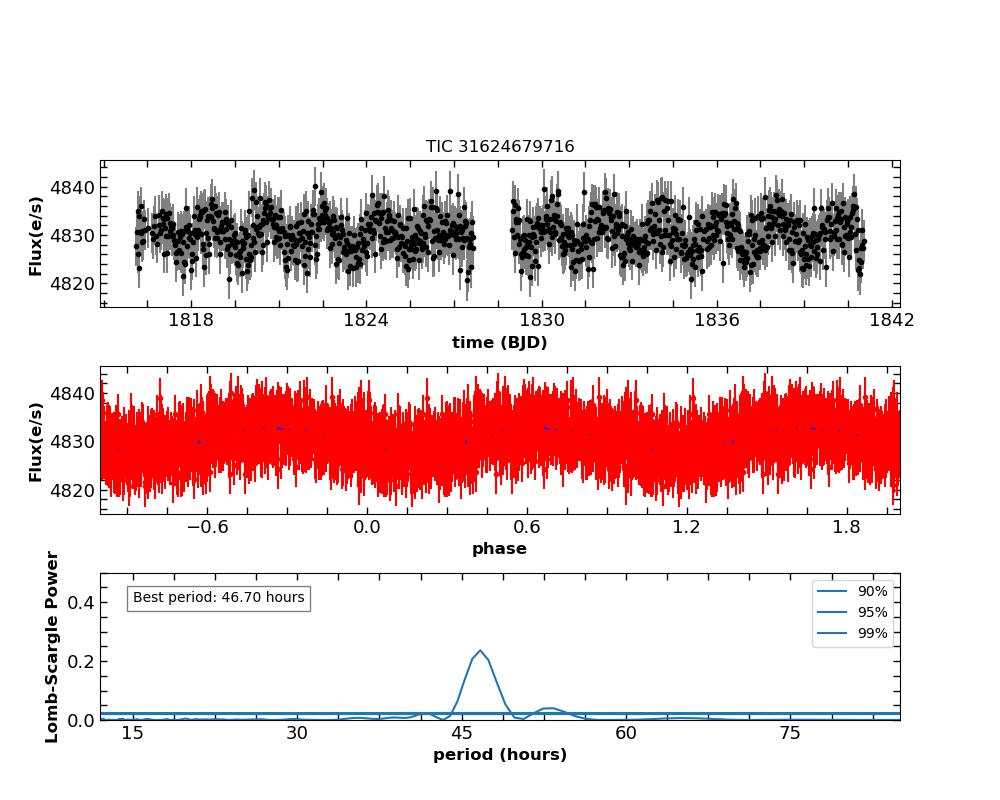}
	\caption{Light curves for the star TIC 31624679716. Upper panel shows light curve, while middle and lower panels show phase-folded light curve and the periodogram.}
        \label{TIC_31624679716}
\end{figure}

The light curve of the star TIC 31624679716 $(RA=51.8077 deg, DEC=56.3553 deg)$ is shown in Figure \ref{TIC_31624679716}. The upper panel shows the light curve, while the middle panel shows the phase folded light curve. The light curve shows a coherent periodic variability in the star. The change in magnitude in the light curve is $\sim$ 0.02 mag. The lower panel shows the Lomb-Scargle periodogram computed from the data shown in the upper panel. The detected period of the variable star is found to be 46.70 hrs. 
\subsection{TIC 31632068918}
\begin{figure}\centering
\centering
	\includegraphics[width=0.65\textwidth]{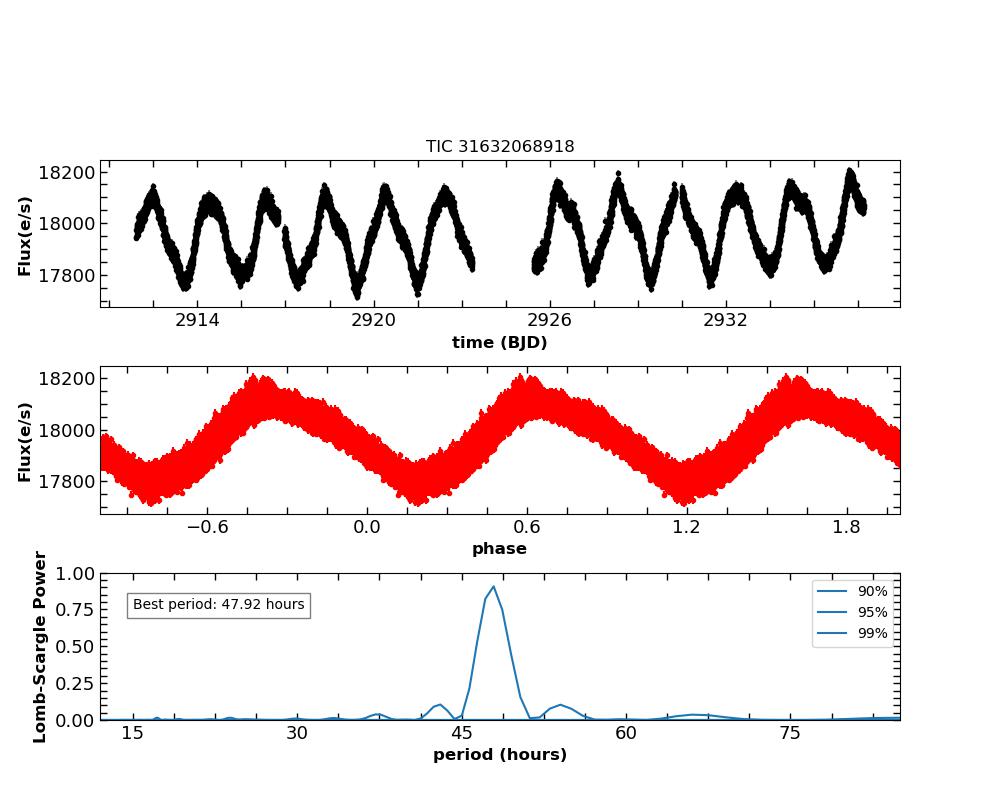}
	\caption{Same as Figure \ref{TIC_31624679716} for the variable star TIC 31632068918.}
        \label{TIC_31632068918}
\end{figure}

Figure \ref{TIC_31632068918} shows the light curve of the variable star TIC 31632068918 $(RA=52.0315 deg, DEC=56.3031 deg)$. The upper and middle panels display the light curve and phase folded light curve, while the lower panel shows the Lomb-Scargle periodogram computed from the data shown in the upper panel. The variation in magnitude is estimated to be $\sim$ 0.01 mag. The derived period of this variable star is 47.92 hrs.

\subsection{TIC 3163212399}
\begin{figure}\centering
\centering
	\includegraphics[width=0.65\textwidth]{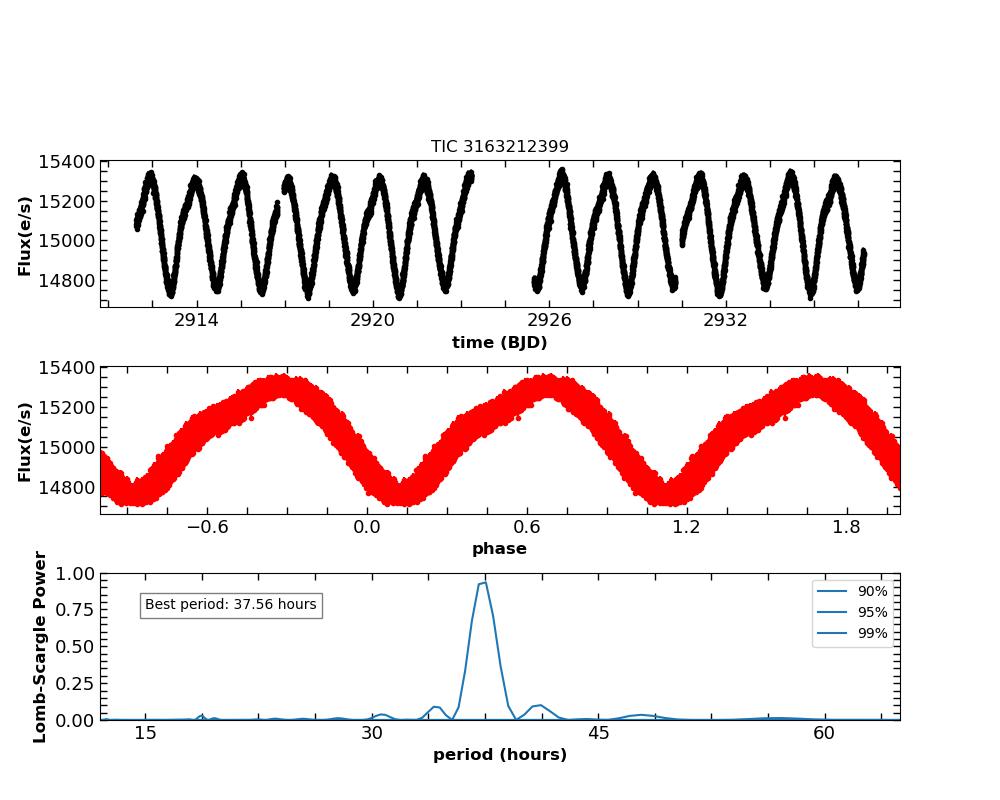}
	\caption{Same as Figure \ref{TIC_31624679716} for the variable star TIC 3163212399.}
        \label{TIC_3163212399}
\end{figure}

The $TESS$ light curve of the variable star TIC 3163212399 $(RA= 52.0408 deg, DEC= 56.5152 deg)$ is shown in Figure \ref{TIC_3163212399}. The upper, middle and lower panels display the light curve, phase folded light curve and Lomb-Scargle periodogram. A coherent periodicity is observed in the light curve, indicating a magnitude variation of $\sim$ 0.04 mag for the star. The determined period of the star is 37.56 hours.

\subsection{H-R Diagram and classification of the variables}
\begin{figure}[H]\centering
\centering
	\includegraphics[width=0.65\textwidth]{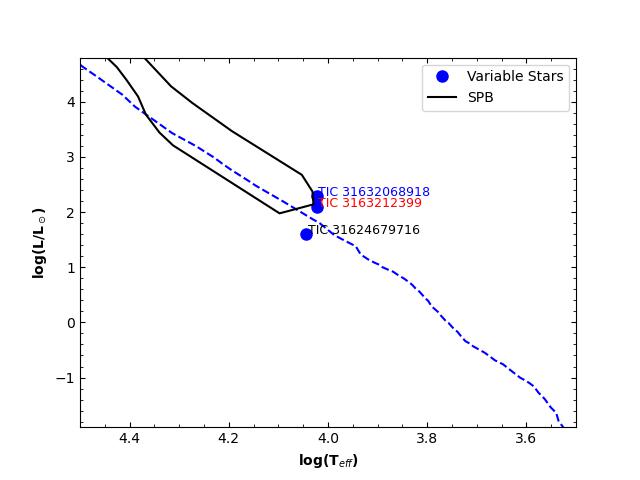}
	\caption{Hertzsprung-Russell (log(L/L$_\odot$) vs. log(T$_{eff}$)) diagram for periodic variables within the cluster region.}
        \label{variable_LF}
\end{figure}

The variable stars can be classified using the location of the stars on the H-R diagram, the shape of their light curve, variability amplitude, and periods. The variable stars searched in this study are plotted in the H-R diagram depicted in Figure \ref{variable_LF}. The dotted line is the main-sequence curve taken from \citet{pecaut2013}, and the black curve is the region for the slow pulsating B-type (SBP) stars \citep{Miglio2007}.  Our study has identified three variables near the SBP region. The period ranges from 0.30 to 5 days for SPB-type stars, and the magnitude variation should be generally less than 0.10 mag. We classify these stars as SPB variables based on their location in the H-R diagram, period, spectral type, magnitude variation, and light curve shape.

\section{Conclusion}
\label{summary}
We have studied the intermediate age open star cluster King 6 using $UBV(RI)_c$, $2MASS$ $JHK$, $GAIA$    
 DR3 and $TESS$ data. The main findings of our analysis are as follows:
\begin{enumerate}
    \item The cluster radius is found to be 9$^\prime$.0, which corresponds to 1.98 pc at the cluster's distance. The tidal radius of the cluster is determined to be 3.18 pc. 
    \item  From the colour-colour diagram, we have estimated $E(B-V)$ = 0.58 $\pm$ 0.03 for the cluster. The $JHK$ data, in combination with the optical data, provide $E(J-K)$ = 0.24 $\pm$ 0.03 while $E(V-K)$ = 1.50 $\pm$ 0.01. The value of $R_v$ is found to be 2.78 $\pm$ 0.30. Hence, our analysis indicates that the interstellar extinction law is normal in the direction of the cluster.
    \item The age of the cluster is found to be 251 Myrs, which is determined using the theoretical isochrone of \citet{Bertelli1994} of solar metallicity $Z = 0.02$.
    \item The distance of the cluster is estimated to be 724 $\pm$ 5 pc. The distance value is supported by the values determined using near-IR data and parallax.
    \item The luminosity function is determined by considering probable members. We found an increasing trend up to $M_v$ $\sim$ 4.0 mag and a dip at $\sim$ 5.50 mag for the cluster. The overall mass function slope is 0.57 $\pm$ 0.28, which is less than the \citet{Salpeter1955} value.
    \item There is clear evidence of mass segregation within the cluster. The Kolmogorov-Smirnov test indicates a 98\% confidence level for this mass segregation. Additionally, the relaxation time of the cluster is determined as 6.0 Myr, suggesting that the cluster is dynamically relaxed.
    
    \item We searched variable stars in the cluster region finding three variables with $TESS$ ID $TIC$ 31624679716, $TIC$ 31632068918, and $TIC$ 3163212399 for the first time. The variation in magnitude of the variable stars is 0.02, 0.01, and 0.04 mag, and the period is 46.70, 47.92, and 37.56 hrs, respectively classifying them as Slow Plusating B-type stars.
\end{enumerate}

\section{Acknowledgement}
\label{sec:acknow}
We are thankful to the anonymous referee for their constructive suggestion on this paper which has improved it a lot. We also acknowledge the use of data from the European Space Agency (ESA) mission Gaia (processed by the Gaia Data Processing and Analysis Consortium, DPAC), which is publicly available from the Gaia Data Release 3 (DR3) and the use of data products from the Two-Micron All Sky Survey ($2MASS$), a joint project of the University of Massachusetts and the Infrared Processing and Analysis Center/California Institute of Technology, funded by NASA and the National Science Foundation. This work has made use of data obtained from the Transiting Exoplanet Survey Satellite ($TESS$) mission, which is funded by NASA  Science Mission Directorate. Furthermore, we are grateful to the Aryabhatta Research Institute of Observational Sciences (ARIES) for its support and contribution to this research.


\end{document}